\documentclass[aps,prb,showpacs,showkeys]{revtex4}

\newcommand{\beq}{\begin{equation}}
\newcommand{\eeq}{\end{equation}}
  \newcommand{\beql}[1]{\begin{equation}\label{eq:#1}}
  \newcommand{\beqa}{\begin{eqnarray}}
  \newcommand{\eeqa}{\end{eqnarray}}
  \newcommand{\beqas}{\begin{eqnarray*}}
  \newcommand{\eeqas}{\end{eqnarray*}}
  \newcommand{\qed}{{\em QED}}
  
  \newcommand*{\Q}{\mathbf{Q}}
  \newcommand*{\R}{\mathbf{R}}

  \newcommand*{\bA}{\mathbf{A}}

  \newcommand*{\bP}{\mathbf{P}}
 \newcommand*{\bQ}{\mathbf{Q}}
  
  \newcommand*{\bS}{\mathbf{S}}

  \newcommand*{\bx}{\mathbf{x}}
  
  \newcommand*{\cA}{\mathcal{A}}
  \newcommand*{\cB}{\mathcal{B}}

  \newcommand*{\cE}{\mathcal{E}}

  \newcommand*{\cH}{\mathcal{H}}
  \newcommand*{\cI}{\mathcal{I}}
  
  \newcommand*{\cK}{\mathcal{K}}
  \newcommand*{\cL}{\mathcal{L}}
  \newcommand*{\cM}{\mathcal{M}}

  \newcommand*{\cQ}{\mathcal{Q}}

  \newcommand*{\cW}{\mathcal{W}}

  \newcommand*{\tA}{\tilde{A}}
  \newcommand*{\tB}{\tilde{B}}

  \newcommand*{\tL}{\tilde{L}}

  \newcommand*{\tS}{\tilde{S}}

  \newcommand*{\al}{\alpha}

  \newcommand*{\da}{\dagger}
  
  \newcommand*{\ep}{\epsilon}
  \newcommand*{\et}{\eta}

  \newcommand*{\nn}{\nonumber}

  \newcommand*{\ps}{\psi} 
  \newcommand*{\rh}{\rho}
  \newcommand*{\si}{\sigma} 
  \newcommand*{\ta}{\tau}

  \newcommand*{\De}{\Delta}                                          
  
  \newcommand*{\Eq}[1]{Eq.~(\ref{eq:#1})}
  \newcommand*{\Ga}{\Gamma}

  \newcommand*{\Tr}{\mbox{\rm Tr}}

  \newcommand*{\eq}[1]{(\ref{eq:#1})}
\newcommand*{\bold}{\mathbf}
\newcommand*{\bra}[1]{\langle#1|}
\newcommand*{\ket}[1]{|#1\rangle}

\newcommand*{\bracket}[1]{\langle#1\rangle}

\newtheorem{Theorem}{Theorem}[section]

\newenvironment{Proof}{\begin{trivlist}
 \item[\hskip \labelsep {\em \indent Proof.}]}{\qed\end{trivlist}}
\newcommand{\dom}{{\rm dom}}

\newcommand{\tc}{\tau c}

\newcommand{\X}{\mathbf{R}^{d}}

\begin{document}
\title{Uncertainty Principle for Quantum Instruments and Computing}
\author{Masanao Ozawa}
\email{ozawa@mailaps.org}
\affiliation{Graduate School of Information Sciences,
T\^{o}hoku University, Aoba-ku, Sendai,  980-8579, Japan}

\begin{abstract}
The notion of quantum instruments is formalized as statistical
equivalence  classes of all the possible quantum measurements and
mathematically characterized  as normalized completely positive
map valued measures under naturally acceptable axioms. Recently,
universally valid uncertainty relations have been established to set 
a precision limit for any instruments given a disturbance constraint 
in a form more general than the one originally proposed by
Heisenberg.   One of them leads to a quantitative  generalization of
the Wigner-Araki-Yanase theorem on the precision limit of
measurements  under conservation laws.    Applying this, a
rigorous lower bound is obtained for the gate  error probability of
physical implementations of Hadamard gates  on a standard qubit
of a spin 1/2 system by interactions  with control fields or ancilla
systems obeying the angular  momentum conservation law.
\end{abstract}
\maketitle

\section{Introduction}

Heisenberg's uncertainty principle\cite{Hei27} in its original
formulation has been understood to set a limitation on
measurements by asserting a lower bound of the product of the
imprecision  of measuring one observable and the disturbance
caused in another noncommuting  observable. However, the
mathematical formulation established by 
Kennard,\cite{Ken27} Robertson,\cite{Rob29} and 
Heisenberg\cite{Hei30} merely
represents the trade-off between standard deviations
of noncommuting observables in a given state and 
does neither allow such an  interpretation, nor has served to
provide a universally valid precision limit of
measurements.   Although such a state of the art has been
undoubtedly resulted from the lack of reliable general
measurement theory, the recent development of the theory 
has made possible to establish desirable operational uncertainty 
relations universally valid for the most general class of quantum
measurements, which will be useful  for precision measurements,  quantum
information, and quantum computing. 
This paper reports the development on the operational uncertainty
relations, and their applications to operational decoherence of
quantum logic gates based on the authors recent
work.\cite{02CLU}$^{\mbox{--}}${\cite{q-ph/0310070}}

\section{Quantum Instruments}

Since von Neumann's axiomatization\cite{vN32} of quantum mechanics, 
we have definite answers to questions as to  
what are general states and what are general observables. 
However, the question was left unanswered for long time
as to what are general measurements.
Towards this problem, Davies and Lewis\cite{DL70} (DL) 
introduced the mathematical formulation of the 
notion of  ``instrument'' as normalized positive map valued measures
(DL instruments),
and showed that this notion generally describes the  
statistical properties of measurement,
so that joint
probability distributions of any sequence of measurements  are
determined by their corresponding DL instruments.
However, the question was left open for some time
as to whether every DL instrument corresponds to a possible 
measuring apparatus.\cite{Yue87}
In order to solve this question, 
Refs.~\onlinecite{83CR,84QC} introduced a general class of mathematical
models of measuring processes (indirect measurement models)
and showed that the statistical properties 
given by any such model is described by a normalized completely 
positive map valued measure (CP instrument),
and conversely that any CP instrument arises in this way.  
Thus, we can naturally conclude that measurements are represented by
CP instruments, just as states are represented by density operators and
observables are represented by self-adjoint operators.

Ref.~\onlinecite{03URN} introduced the notion of statistical equivalence
of measurements so that two measuring apparatuses are statistically
equivalent if and only if they are interchangeable without 
affecting joint probability distributions of any sequences 
of measurements, 
and reformulated the above characterization of measurements 
under the following naturally acceptable axioms.

(i) {\em Mixing law}: If two apparatuses
are applied to a single system in succession, the joint probability 
distribution of outputs from those two  apparatuses 
depends affinely on the input state.

(ii) {\em Extendability axiom}: Every apparatus measuring one
system can be trivially extended to an apparatus measuring
a larger system including the original system without
changing the statistics. 

(iii) {\em Realizability postulate}:  Every indirect measurement
model corresponds to an apparatus whose measuring process
is described by that model.

Under the above axioms (i)--(iii), it was proven in Ref.~\onlinecite{03URN}
that the statistical equivalence classes of apparatuses are in one-to-one 
correspondence with the CP instruments.
Thus, we established the notion of  ``instrument'' as the function
of a measuring apparatus by the mathematical notion
``CP instrument'' that represents the statistical equivalence 
class of a measuring apparatus.
In this paper, we shall thus define ``instruments'' as CP instruments.

Let $\cH$ be a Hilbert space.  
A map $\Pi:\cB(\R^d) \to \cL(\cH)$ is called a 
{\em probability operator valued measure} (POVM) for $(\cH,\R^{d})$,
where $\cB(\R^d)$ stands for the Borel $\si$-field of the Euclidean space
$\R^d$ and $\cL(\cH)$ stands for the space of 
bounded linear operators on $\cH$,
if it satisfies the following conditions:
(i) For any disjoint sequence $\De_{1}, \De_{2},\ldots$ in
$\cB(\R^d)$, we have
$ 
\Pi(\bigcup_{i=1}^{\infty}\De_i) = \sum_{i=1}^{\infty} \Pi(\De_i),
$ 
where the sum is convergent in the weak operator topology.
(ii) $\Pi(\R^{d}) = I$.

A linear transformation $T:\ta c(\cH) \to \ta c(\cH)$ is called {\em
completely positive\/} (CP), where $\tc(\cH)$ stands for the space 
of trace class operators on $\cH$, if $T\otimes id$ on
$\tc(\cH\otimes\cH)$ is a positive map, where $T\otimes id$ is
the extension of $T$ to $\tc(\cH\otimes\cH)$ determined by
$(T\otimes id)(\rh\otimes\si)=(T\rh)\otimes\si$ for elementary
tensors.
We shall denote the space of all CP maps on $\ta c(\cH)$
by ${\rm CP}_{\cH}$.  
The {\em dual} $T^{*}:{\cal L}({\cal H}) \to \cL({\cal H})$ 
of $T \in {\rm CP}_{\cH}$ is defined by the relation
$\mbox{\rm Tr}[T^{*}(a)\rho ] = \mbox{\rm Tr}[aT(\rho )]
$ for all $a \in  {\cal L}({\cal H})$ and $\rho  \in  \ta c({\cal H})$. 
Then, $T^{*}$ is a normal CP map on $\cL(\cH)$,
and $T$ is trace-preserving if and only if $T$ is
unit-preserving.\cite{Dav76}

A map $\cI: \cB(\R^d) \to {\rm CP}_{\cH}$
is  called an {\em instrument\/} for $(\cH,\R^{d})$,
if it satisfies the following 
conditions:\cite{DL70,84QC}

(i) For any disjoint sequence $\De_{1}, \De_{2},\ldots$ in $\cB(\R^d)$,
\beqa
     \cI(\bigcup _{i=1}^{\infty} \De_i) 
= \sum _{i=1}^{\infty} \cI(\De_i),
\eeqa
\sloppy where the sum is convergent in the strong operator topology
of ${\rm CP}_{\cH}$.

(ii) For any $\rho  \in  {\ta c}({\cal H})$,
\beqa 
      \mbox{\rm Tr}[\cI({\bold R}^d )\rho ] = \mbox{\rm Tr}[\rho ].
\eeqa 

For any instrument $\cI$, the relation
\beqa\label{eq:POVMofI}
\Pi(\De) = \cI(\De)^{*}I
\eeqa\noindent
for any $\De \in \cB(\R^d)$ determines a POVM $\Pi$, called the
{\em POVM} of $\cI$.   Conversely, it is known\cite{84QC} that every
POVM $\Pi$ has at least one instrument $\cI$ satisfying  \Eq{POVMofI}.
For any state $\rh$, the relation
\beqa
\mu(\De)=\Tr[\cI(\De)\rh]=\Tr[\Pi(\De)\rh]
\eeqa
defines a probability measure on $\cB(\X)$, which called the {\em
output probability distribution} of  $\cI$ in $\rh$.
Let $\De\in\cB(\X)$.  The CP map $\cI(\De)$ is
called the {\em operation} of $\cI$ {\em given} $\De$,
and $\cI(\X)$ is called the {\em nonselective operation} of 
$\cI$.  For any Borel set $\De\in\cB(\X)$ and state $\rh$
with $\Tr[\cI(\De)\rh]>0$,
the state 
\beqa
\rh_{\De}=\frac{\cI(\De)\rh}{\Tr[\cI(\De)\rh]}
\eeqa 
is called the {\em output state} of $\cI$ for the {\em input state} $\rh$
{\em given} $\De$.

\sloppy
A finite set $\{A_1,\ldots,A_n\}$ of observables are called 
{\em compatible}, 
if
\beqa 
[E^{A_i}(\De_1),E^{A_j}(\De_2)] = 0
\eeqa
for all $i,\ j = 1,\ldots,n$ and 
$\De_1,\ \De_2 \in \cB(\R)$, where $E^{A_i}$ stands for the 
spectral measure corresponding to $A_{j}$.
In this case, we shall write $[A_{i},A_{j}]=0$ for any $i,j$.

In this paper, by a {\em measuring process}
we shall generally mean an
experiment described as follows.  
Let $\bP$ be a quantum system, called a {\em probe system}, described by
a Hilbert space $\cK$.  The system $\bP$ is coupled to the system 
$\bS$ during a finite time interval $(t,t+\De t)$.
Denote by $U$ the unitary operator on $\cH\otimes\cK$ corresponding
to the time evolution of the system $\bS+\bP$ for the time interval
$(t,t+\De t)$.
At time $t$, the time of measurement, 
the probe system $\bP$ is prepared in
a fixed state $\si$. 
At time $t+\De t$, the time just after the measuring interaction,
the systems $\bS$
and $\bP$ are separated and
 a  compatible observables $M_1,\ldots,M_d$ of
the system $\bP$, called the {\em meter observables}, 
are measured precisely. 
Thus, any measuring process is characterized by 
a (3+d)-tuple $\cM = (\cK,\si,U,M_1,\ldots,M_d)$ 
consisting of a Hilbert space $\cK$, 
a density operator $\si$ on $\cK$, a unitary 
operator $U$ on $\cH\otimes\cK$ and a compatible sequence
$(M_1,\ldots,M_d)$ of self-adjoint operators on $\cK$.
Every measuring process
$\cM = ({\cal K},\sigma ,U,M_1,\ldots,M_d)$ determines a
unique instrument
$\cI:\cB(\R^d)\to {\rm CP}_{\cH}$, called the {\em instrument} of $\cM$, 
by the following relation
\beqa\label{(cpom)}
\cI(\De_1\times\cdots\times\De_d)\rh 
= \Tr_{\cK}\left\{\left[1\otimes E^{M_1}(\De_1)\cdots E^{M_d}(\De_d)\right]
U(\rh\otimes \si)U^{\dag}\right\},
\eeqa
for all $\rh \in \ta c(\cH)$ and $\De_1,\ldots,\De_d \in \cB(\R)$, where
${\Tr}_{\cK}$ stands for the partial trace operation of $\cK$.

Given an instrument $\cI$ for $(\cH,\R^{d})$, 
any measuring process
$\cM = ({\cal K},\sigma ,U,M_1,\ldots,M_d)$ which satisfies
Eq.~(\ref{(cpom)}) is called a {\em realization} of $\cI$.  
An instrument is called {\em realizable} if it has at least
one realization.  The definition of instruments are justified
by the following theorem.\cite{83CR,84QC}

\begin{Theorem}
Every instrument $\cI:\cB(\R^d) \to {\rm CP}_{\cH}$ has at
least one realization 
$\cM=({\cal K},\sigma ,U,M_1,\ldots,M_d)$ such that $\si$
is a pure state.
\end{Theorem}

The following theorems\cite{83CR,84QC} are immediate consequences
from the above theorem; Theorem \ref{th:real-operation} 
was also obtained
by Kraus\cite{Kra83}  independently.

\begin{Theorem}\label{th:realPOVM}
For any POVM $\Pi:\cB(\R^d) \to \cL(\cH)$, there exists a measuring
process $\cM=(\cK,\ket{\xi}\bra{\xi},U,M_1,\ldots,M_d)$ satisfying
the relation
\beqa\label{eq:realPOVM}
\Pi(\De_1\times\cdots\times\De_d) 
= \bracket{\xi|U^{\dag}[I\otimes
M_1(\De_1)\cdots M_d(\De_d)]U|\xi}
\eeqa
for all $\De_1,\ldots,\De_d \in \cB(\R^d)$, where
$\bracket{\xi|\cdots|\xi}$ stands for the partial inner product
such that 
$\bracket{\ps|\bracket{\xi|\cdots|\xi}|\ps}
=\bracket{\ps\otimes\xi|\cdots| \ps\otimes\xi}$.
\end{Theorem}
\begin{Theorem}\label{th:real-operation}
For any trace preserving CP map $T:\tc(\cH)\to\tc(\cH)$, 
there exist a Hilbert space $\cK$, a unit vector $\xi\in\cK$, and
a unitary operator $U$ on $\cH\otimes\cK$ satisfying the relation
\beqa
T\rh=\Tr_{\cK}[U(\rh\otimes\ket{\xi}\bra{\xi})U^{\da}]
\eeqa
for all $\rh\in\tc(\cH)$.
\end{Theorem}
\begin{Theorem}
For any normal unit preserving CP map $T:\cL(\cH)\to\cL(\cH)$, 
there are a Hilbert space $\cK$, a unit vector $\xi\in\cK$, and
a unitary operator $U$ on $\cH\otimes\cK$ satisfying the relation
\beqa
T a=\bracket{\xi|U^{\da}(a\otimes I)U|\xi}
\eeqa
for all $a\in\cL(\cH)$.
 \end{Theorem}
 
Let $\Pi$ be a POVM for $(\cH,\R)$.
Let $f(x)$ be a real Borel function on $\R$. 
Denote by $\int f(x) d\Pi(x)$, or $\int f d\Pi$ for short,
the symmetric operator defined by 
\begin{equation}
\bracket{\xi|\int f(x) d\Pi(x)|\et} 
= \int_\R f(x)\,d\bracket{\xi|\Pi (x)|\et}
\end{equation}
for any $\xi,\et\in\dom(\int f(x)d\Pi(x))$, where the domain is defined by
\beqa
\dom\left(\int f(x) d\Pi(x)\right)
&=& \left\{\xi\in\cH \mid \int_\R f(x)^2 \,d\bracket{\xi|\Pi (x)|\xi} <
\infty
\right\}.\quad
\eeqa
The {\em n-th moment operator of $\Pi $}, denoted by 
 $O^{(n)}(\Pi )$, is defined by
\beqa
O^{(n)}(\Pi )=\int_{\R}x^{n}\, d\Pi(x).\label{eq:moment2}
\eeqa
We shall write $O(\Pi)=O^{(1)}(\Pi)$.
The {\em mean} $\bracket{O(\Pi)}$ and 
the {\em standard deviation} $\De(\Pi)$ of 
POVM $\Pi$ is given, if the integral converges, by
\beqa\label{(3.a4)}
\bracket{O(\Pi)} &=& \Tr[O(\Pi)\rh],       \\
\De(\Pi) &=& (\bracket{O^{(2)}(\Pi)} -
\bracket{O(\Pi)}^2)^{1/2}.                     
\eeqa
For any observable $A$, we have $A=O(E^{A})$ and
the mean of $A$ in state $\rh$ defined by $\bracket{A}=
\Tr[A\rh]$ satisfies $\bracket{A}=\bracket{O(E^{A})}$.
The standard deviation of $A$ in state $\rh$ defined by $\De
A=(\bracket{A^{2}}-\bracket{A}^{2})^{1/2}$ satisfies
$\De A=\De(E^{A})$.
By the Robertson uncertainty relation,\cite{Rob29} for any
state 
$\rh$ and any pair of observables $A$, $B$ with $\De A$, 
$\De B<\infty$, we have
\beqa\label{eq:Robertson}
\De A\,\De B\ge \frac{1}{2}\left|\bracket{[A,B]}\right|,
\eeqa
where $\bracket{[A,B]}=\Tr\{[A,B]\rh\}$.

\section{Operational Uncertainty Relations}

In this section,
we generalize Heisenberg's noise-disturbance uncertainty 
relation to a relation that holds for any
instruments,  from which
conditions are obtained for measuring instruments to satisfy 
Heisenberg's relation.\cite{03UVR,03HUR,03URN} 
In particular, every instrument with the noise and the disturbance
uncorrelated with the measured object is proven
to satisfy Heisenberg's relation.\cite{03URN}

\subsection{Uncertainty relations for joint direct measurements}

Let $\cW$ be a Hilbert space and $\si$ be a density operator on $\cW$.
Let $A,C$ be two observables on $\cW$.   The noise operator
$N(C,A)$ for $C$ in measuring $A$ and the root-mean-square
noise
$\ep(C,A,\si)$ for $C$ in measuring $A$ in $\si$ are defined by
\beqa
N(A,C)&=&C-A,\\
\ep(A,C,\si)&=&\Tr[N(A,C)^{2}\si]^{1/2}.
\eeqa
Under the above definitions, we have the following.\cite{03UVR}
\begin{Theorem}
\label{th:model}
For any four observables $A,B,C,D$ on $\cW$,  
if $C$ and $D$ are commuting, we have
\beqa
& &\ep( A)\,\ep( B)
+\ep( A)\,\De B+\De A\,\ep( B)\nn\\
&\ge &\De N_{A}\,\De N_{B}
+\De N_{A}\,\De B+\De A\,\De N_{B}\label{eq:1-a}\\
&\ge&\De N_{A}\,\De N_{B}+
\frac{1}{2}|\bracket{[N_{A},B]}|
+\frac{1}{2}|\bracket{[A,N_{B}]}|\label{eq:1-b}\\
&\ge&\frac{1}{2}|\bracket{[A,B]}|\label{eq:1-c}
\eeqa
for any state $\si$ for which all the relevant terms are
finite,
where
$N_{A}=N(A,C)$, $N_{B}=N(B,D)$,
$\ep(A)=\ep(A,C,\si)$,  $\ep(B)=\ep(B,D,\si)$,
$\De$ stands for the standard deviation in $\si$,
and $\bracket{\cdots}$ stands for the mean value in $\si$.
\end{Theorem}
\begin{Proof}
By definition, we have
\begin{eqnarray}
C&=&A+N_{A},\label{eq:NO}\\
D&=&B+N_{B}.\label{eq:DO}
\end{eqnarray}
From $[C,D]=0$, we have
the following commutation relation for noise operators,
\beqa
[N_{A},N_{B}]+[N_{A},B]+[A,N_{B}]=-[A,B].
\eeqa
Taking the modulus of means of the both sides
 and applying the triangular inequality, we have
\begin{eqnarray}
|\bracket{[N_{A},N_{B}]}|+|\bracket{[N_{A},B]}|
+|\bracket{[A,N_{B}]}|
\ge|\bracket{[A,B]}|.
\end{eqnarray}
By Robertson's inequality,  \Eq{Robertson}, we have
\beqa
\De N_{A}\De N_{B}&\ge&\frac{1}{2}|\bracket{[N_{A},N_{B}]}|,\\
\De {A}\De N_{B}&\ge&\frac{1}{2}|\bracket{[{A},N_{B}]}|,\\
\De N_{A}\De {B}&\ge&\frac{1}{2}|\bracket{[N_{A},{B}]}|,
\eeqa
so that inequalities \eq{1-b} and \eq{1-c} follow. 
Since the variance is not greater than the mean square,  we have 
\begin{eqnarray}
\ep( A)&\ge&\De N_{A},\\
\ep( B)&\ge&\De N_{B},
\end{eqnarray}
and hence inequality \eq{1-a} follows.
\end{Proof}

\subsection{Uncertainty relations for joint indirect measurements}

Let $\cH$ be a Hilbert space and $\rh$ be a density operator on $\cH$.
Let $A$ be an observable on $\cH$ 
and let $\Pi^{A}$ be any POVM on $\cH$.
The mean noise operator $n(A,\Pi^{A})$ for $\Pi^{A}$ in
measuring $A$, 
the mean noise noise $\bar{n}(A,\Pi^{A})$,
the root-mean-square noise $\ep(\Pi^{A},A,\rh)$,
and 
the standard deviation $\De N(A,\Pi^{A})$ of the noise
for $\Pi^{A}$ in measuring $A$ in $\rh$
 are defined by
\beqa
n(A,\Pi^{A})&=&O(\Pi^{A})-A,\\
\bar{n}(A,\Pi^{A},\rh)&=&\bracket{n(A,\Pi^{A})},\\
\ep(A,\Pi^{A},\rh)&=&\bracket{O^{(2)}(\Pi^{A})-O(\Pi^{A})A-AO(\Pi^{A})
+A^{2}}^{1/2},\\
\De N(A,\Pi^{A},\rh)
&=&[\ep(A,\Pi^{A},\rh)^{2}-\bar{n}(A,\Pi^{A})^{2}]^{1/2},
\eeqa
where $\bracket{\cdots}$ stands for the mean value in
the state $\rh$, i.e., $\bracket{\cdots}=\Tr[\cdots\ \rh]$.

By the Naimark theorem,\cite{RN55}
there exist a Hilbert space
$\cW$, an isometry $V:\cH\to\cW$,
and a self-adjoint operator $C$ such that 
\beqa\label{eq:Naimark}
\Pi^{A} (\De)=V^{\da}E^{C}(\De)V
\eeqa 
for every Borel set $\De$.
We shall call any triple $(\cW,V,C)$ satisfying \Eq{Naimark}
a {\em Naimark extension}  of $\Pi^{A} $. 
Then, we have the following.\cite{03URN}

\begin{Theorem}
\label{th:Naimark}
For any Naimark extension $(\cW,V,C)$ of a POVM $\Pi^{A} $ on $\cH$,
we have
\beqa
n(A,\Pi^{A})&=&V^{\da}CV-A,\\
\bar{n}(A,\Pi^{A})&=&\Tr[(V^{\da}CV-A)\rh],\\
\ep(A,\Pi^{A},\rh)
&=&\|CV\sqrt{\rh}-VA\sqrt{\rh}\|_{HS},\\
\De N(A,\Pi^{A},\rh)&=&\|CV\sqrt{\rh}-VA\sqrt{\rh}-
\bar{n}(A,\Pi^{A})\sqrt{\rh}\|_{HS},
\eeqa
where $\|\cdots\|_{HS}$ stands for the Hilbert-Schmidt norm.
\end{Theorem}

The following theorem\cite{03URN} characterizes POVMs with 
zero-noise.

\begin{Theorem}\label{th:zero-distance-1}
For any POVM $\Pi^{A} $ on $\cH$ and any observable $A$ on $\cH$,
the following conditions are equivalent.

(i) $\Pi^{A} =E^{A}$.

(ii) $\ep(A,\Pi^{A},\rh)=0$ for any state $\rh$.

(iii) $\ep(A,\Pi^{A},\rh)=0$ for a faithful state $\rh$.

(iv) $\ep(A,\Pi^{A},\ket{n})=0$ for any $\ket{n}$ in 
an orthonormal basis $\{\ket{n}\}$.

(v) $\ep(A,\Pi^{A},\ps)=0$ for any state vector $\ps\in\cH$.

\end{Theorem}

We call any POVM for $(\cH,\R^{2})$ the {\em joint POVM}
for $\cH$.
The {\em marginal POVMs} $(\Pi^{A},\Pi^{B})$ of joint POVM
$\Pi$ are defined by
$\Pi^{A}(\De)=\Pi(\De\times\R)$ and 
$\Pi^{B}(\Ga)=\Pi(\R\times\Ga)$ for any $\De,\Ga\in\cB(\R)$.

Under the above definitions, we have the following.\cite{03UVR}
\begin{Theorem}
\label{th:POVM}
For any two observables $A,B$ on $\cH$,
and joint POVM $\Pi$ for $\cH$ with marginal POVMs
 $(\Pi^{A},\Pi^{B})$,
we have
\beqa
& &\ep( A)\,\ep( B)
+\ep( A)\,\De B+\De A\,\ep( B)\nn\\
&\ge &\De N_{A}\,\De N_{B}
+\De N_{A}\,\De B+\De A\,\De N_{B}\label{eq:2-a}\\
&\ge&\De N_{A}\,\De N_{B}+
\frac{1}{2}|\bracket{[n_{A},B]}|
+\frac{1}{2}|\bracket{[A,n_{B}]}|\label{eq:2-b}\\
&\ge&\frac{1}{2}|\bracket{[A,B]}|\label{eq:2-c}
\eeqa
for any state $\rh$ for which all the relevant terms are
finite, where
$n_{A}=n(A,\Pi^{A})$, $n_{B}=n(B,\Pi^{B})$, 
$\ep(A)=\ep(A,\Pi^{A},\rh)$,  $\ep(B)=\ep(B,\Pi^{B},\rh)$,
$\De N_{A}=\De N(A,\Pi^{A},\rh)$,
$\De N_{B}=\De N(B,\Pi^{B},\rh)$,
while $\De A,\De B$ stand for the standard deviations in $\rh$,
and $\bracket{\cdots}$ stands for the mean value in $\rh$.
\end{Theorem}
\begin{Proof}
Let $(\cK,\xi,U,M_{1},M_{2})$ be a realization of $\Pi$ given
in Theorem \ref{th:realPOVM}.
By defining $C=U^{\da}(I\otimes M_{1})U$ 
and $D=U^{\da}(I\otimes M_{2})U$ in \Eq{realPOVM}, 
we have commuting observables $C,D$ on $\cH\otimes\cK$ 
such that
$
\Pi(\De\times\Ga)=\bracket{\xi|E^{C}(\De)E^{D}(\Ga)|{\xi}}
$
for any $\De,\Ga\in\cB(\R)$.
Then, from Theorem \ref{th:Naimark} we have
\beqa
n(A,\Pi^{A})&=&\bracket{\xi|N(\tA,C)|\xi},\\
\ep(A,\Pi^{A},\rh)&=&\ep(\tA,C,\rh\otimes\ket{\xi}\bra{\xi}),\\
\De A&=&\De \tA\\
\De N(A,\Pi)&=&\De N(\tA,C)
\eeqa
and analogous relations for $B$ and $D$.
By the relations
\beqas
\bracket{N(\tA,C)\tB}
&=&\Tr\{N(\tA,C)[(B\rh)\otimes\ket{\xi}\bra{\xi}]\}\\
&=&\Tr\{\Tr_{\cK}[N(\tA,C)(I\otimes\ket{\xi}\bra{\xi})](B\rh)\}\\
&=&\Tr[\bracket{\xi|N(\tA,C)|\xi}B\rh]\\
&=&\bracket{n(A,\Pi^{A})B},
\eeqas
 we have
\begin{equation}\label{eq:3618a}
\bracket{[N(\tA,C),{\tB}]}=\bracket{[n(A,\Pi^{A}),B]}.
\end{equation}
Similarly, we also have
\begin{equation}\label{eq:3618b}
\bracket{[{\tA},N({\tB},D)]}=\bracket{[A,n(B,\Pi^{B})]}.
\end{equation}
Therefore, by substituting the above relations, the assertion
follows from Theorem \ref{th:model}.
\end{Proof}

From the above, if $\Pi$ precisely measures $A$, i.e., $\ep(A)=0$, 
we have
\begin{equation}\label{eq:noiseless}
\De A\,\ep(B)\ge\frac{1}{2}|\bracket{[A,B]}|.
\end{equation}

We say that   POVM $\Pi^{A}$ has 
{\em uncorrelated noise} for $A$, 
if the mean noise $\bar{n}(A,\Pi,\rh)$
does not depend on the input state $\rh$, or equivalently, if the mean noise
operator $n(A,\Pi)$ is a constant operator, i.e., $n(A,\Pi)=r I$ for some
$r\in\R$.  We say that POVM $\Pi^{A}$ makes an {\em unbiased
measurement} of  $A$, if $n(A,\Pi)=0$,
so that if $\Pi$ makes an unbiased measurement of  $A$, then
$\Pi$ has uncorrelated noise for $A$.
For the $B$ measurement, the corresponding definitions on uncorrelated noise
and unbiased measurements are introduced analogously.

The relations $n_{A}=r I$ and $n_{B}=r' I$  obviously imply
$[n_{A},B]=[A,n_{B}]=0$, and hence from 
Theorem \ref{th:POVM} we conclude 
the following.\cite{q-ph/0310070}

\begin{Theorem}
If the marginal observables $(\Pi^{A},\Pi^{B})$
of a joint POVM $\Pi$ have uncorrelated noises for $A$ and $B$, respectively,
then we have
\begin{equation}\label{eq:Heisenberg-NAB}
\ep(A)\ep(B)\ge
\De N_{A}\De N_{B}\ge\frac{1}{2}|\bracket{[A,B]}|
\end{equation}
for any state $\rh$.
\end{Theorem}

The above relations were previously proven for the unbiased case 
in Refs.~\onlinecite{91QU,Ish91}.

If $\Pi$ has uncorrelated noise for both $A$
and $B$, we have 
\begin{eqnarray}
\De(\Pi^{A})^{2}&=&(\De A)^{2}+(\De N_{A})^{2}\ge 2\De
A\,\De N_{A},
\\
\De(\Pi^{B})^{2}&=&(\De B)^{2}+(\De N_{B})^{2}\ge
2\De B\,\De N_{B},
\end{eqnarray}
and hence apply \Eq{Robertson} and \Eq{Heisenberg-NAB}
to the product of the above two inequalities, 
we have
\begin{equation}\label{eq:030819d}
\De(\Pi^{A})\De(\Pi^{B})\ge|\bracket{[A,B]}|.
\end{equation}
The above relation has been previously proven for the unbiased case 
in Ref.~\onlinecite{AG88}.

\subsection{Uncertainty relations for instruments}

Let $\cH$ be a Hilbert space
and $\rh$ be a density operator on $\cH$.
Let $B$ be an observables on $\cH$.
Let $T$ be a trace-preserving operation for $\cH$.
The POVM $T^{*}E^{B}$ is defined by 
\beqa
(T^{*}E^{B})(\De)=T^{*}[E^{B}(\De)].
\eeqa
We have $T^{*}(B^{n})=O^{(n)}(T^{*}E^{B})$, if $B$
is bounded. 
The mean disturbance operator $d(B,T)$ of $B$
for $T$,  the root-mean-square disturbance
$\et(B,T,\rh)$ of $B$ for $T$ in $\rh$,
and the standard deviation $\De D(B,T,\rh)$
of the disturbance of $B$ for $T$ in $\rh$
 are defined by
\beqa
d(B,T)&=&n(B,T^{*}E^{B}),\\
\et(B,T,\rh)&=&\ep(B,T^{*}E^{B}),\\
\De D(B,T,\rh)&=&\De N(B,T^{*}E^{B},\rh).
\eeqa
Under the above definitions, we have the following
universal noise-disturbance uncertainty relations.\cite{03UVR}
\begin{Theorem}
\label{th:instruments}
Let $A,B$ be two observables on $\cH$.
For any instrument $\cI$ with POVM $\Pi$  and 
nonselective operation $T$, we have
\beqa
& &\ep( A)\,\et( B)
+\ep( A)\,\De B+\De A\,\et( B)\nn\\
&\ge &\De N_{A}\,\De D_{B}
+\De N_{A}\,\De B+\De A\,\De D_{B}\label{eq:3-a}\\
&\ge&\De N_{A}\,\De D_{B}+
\frac{1}{2}|\bracket{[n_{A},B]}|
+\frac{1}{2}|\bracket{[A,d_{B}]}|\label{eq:3-b}\\
&\ge&\frac{1}{2}|\bracket{[A,B]}|\label{eq:3-c}
\eeqa
for any state $\rh$ for which all the relevant terms are
finite, where
$n_{A}=n(A,\Pi)$, $d_{B}=n(B,T)$, 
$\ep(A)=\ep(A,\Pi,\rh)$,  $\et(B)=\ep(B,T,\rh)$,
and $\De D_{B}=\De D(B,T,\rh)$.
\end{Theorem}

Let $\cI$ be an instrument with POVM $\Pi$ and
nonselective operation $T$.
We say that instrument $\cI$ has {\em uncorrelated noise}
for $A$, if the POVM $\Pi$ has uncorrelated noise, i.e., 
$n(A,\Pi)=r I$ for some
$r\in\R$.
We say that instrument $\cI$ has {\em uncorrelated
disturbance} for $B$, 
if the mean disturbance operator $d(B,T)$ is a constant operator, i.e.,
$d(B,T)=r I$ for some $r\in\R$.

We say that an instrument $\cI$ makes an {\em unbiased
measurement} of  $A$, if $n(A,\Pi)=0$ 
and it makes an {\em unbiased
disturbance} of  $B$, if $d(B,T)=0$.

The universal noise-disturbance
uncertainty relations lead to rigorous conditions on
what  instrument satisfies Heisenberg's noise-disturbance
uncertainty relation, as follows.\cite{03URN}
 
\begin{Theorem}\label{th:HUR}
Let $A$ and $B$ be a pair of observables.
An instrument $\cI$ satisfies Heisenberg's
noise-disturbance uncertainty relation, i.e., 
\beqas
\ep(A)\et(B)\ge \frac{1}{2}|\bracket{[A,B]}|
\eeqas
for any state $\rh$ for which all the relevant terms are finite,
if one of the following conditions holds:

(i) The mean noise operator commutes with $B$ and
the mean disturbance operator commutes with $A$,
i.e.,  
\beqa
[n_A,B]&=&0,\\{}
[d_B,A]&=&0.
\eeqa

(ii) The instrument $\cI$ has both uncorrelated noise for $A$
and uncorrelated disturbance for $B$.

(iii) The instrument $\cI$ makes both unbiased measurement of $A$
and unbiased disturbance of $B$.
\end{Theorem}

For the general case, we have the following trade-off relations for precise
$A$ measurements  or $B$-non-disturbing measurements.\cite{03URN}

\begin{Theorem}\label{th:URND}
For any instrument $\cI$ and observables $A$ and $B$,
if $\et(B)=0$, we have
\begin{equation}\label{eq:URND}
\ep(A)\,\De B
\ge\frac{1}{2}|\Tr([A,B]\rh)|
\end{equation}
for any state $\rh$ for which all the relevant terms are finite.
\end{Theorem}

\begin{Theorem}\label{th:URPM}
For any apparatus $\bA(\bx)$ and observables $A$ and $B$,
if $\ep(A)=0$, we have
\begin{equation}
\De A\,\et(B)
\ge\frac{1}{2}|\bracket{[A,B]}|
\end{equation}
for any state $\rh$ for which all the relevant terms are finite.
\end{Theorem}

\section{Wigner-Araki-Yanase Theorem}

Every interaction brings an entanglement in the basis of a conserved
quantity, so that measurements, and any other quantum state
controls such as quantum information processing, are subject to the
decoherence induced by conservation laws.
One of the earliest formulations of this fact was given by 
the Wigner-Araki-Yanase (WAY)
theorem\cite{Wig52}$^{\mbox{--}}$\cite{Yan61}  stating
that any observable which does not commute with an additive
conserved quantity cannot be measured  with absolute precision.

It is natural to expect that the WAY theorem can be
derived by Heisenberg's uncertainty principle.
However, Heisenberg's relation concludes that
if the measurement does not disturb the total momentum,
the position cannot be measured even with finite precision,
despite that we can do with finite or even 
arbitrarily small noise.\cite{91CP,93WA}
Actually, the WAY theorem does not conclude
unmeasurability of any observables, 
but merely sets the accuracy limit of the measurement with size limited 
apparatus in the presence of bounded conserved quantities. 

We show that the above new formulation of the universal
noise-disturbance uncertainty relation
can be used to derive the quantitative expression of
the WAY theorem as follows.\cite{02CLU}

\begin{Theorem}
Let $\cM=(\cK,\si,U,M)$ be an indirect measurement model for $\cH$
and let $\ep(A)$ be the root-mean-square noise for this measurement
in a state $\rh$,
i.e., $\ep(A)=\ep(A,\Pi,\rh)$, where $\Pi(\De)
=\Tr_{\cK}\{U^{\da}[I\otimes E^{M}(\De)]U(I\otimes \si)\}$.
Let $L_{1}$ and $L_{2}$ be a pair of additive conserved 
quantities on $\cH$ and $\cK$, respectively, i.e.,
$[U,\tL_{1}+\tL_{2}]=0$,
where $\tL_{1}=L_{1}\otimes I_{\cK}$ and 
$\tL_{2}=I_{\cH}\otimes L_{2}$. 
Suppose that the meter observable $M$
commutes with the conserved quantity, i.e.,   $[M,L_{2}]=0$.
Then, we have 
\beqa\label{eq:quantitative-WAY}
\ep(A)^{2}
\ge\frac{|\bracket{[A,L_{1}]}|^{2}}
{4(\De L_{1})^{2}+4(\De L_{2})^{2}},
\eeqa
where the mean and standard deviations are taken in the
state $\rh\otimes\si$.
\end{Theorem}
\begin{Proof}
Let $\cW=L^{2}(\R)$ be the Hilbert space
of one-dimensional mass with position $\hat{q}$ and
momentum $\hat{p}$.
Let $\al>0$ be an arbitrary positive number and let
$\xi$ be a state vector in $\cW$ such that 
$\bracket{\xi|\hat{q}^{2}|\xi}<\al^{2}$.
Consider the indirect measurement model
$$
\cM_{0}=[\cW,\ket{\xi}\bra{\xi},
(I_{\cH}\otimes e^{-i M\otimes\hat{p}/\hbar})
(U\otimes I_{\cW}),\hat{q}]
$$
for $\cH\otimes\cK$ and let $\cI_{0}$ be the corresponding instrument
with POVM $\Pi_{0}$ and nonselective operation $T_{0}$.
Then, we have
\beqa
T_{0}^{*}[E^{\tL_{1}+\tL_{2}}(\De)]
=
\Tr_{\cW}
[U_{1}^{\da}U_{2}^{\da}
(E^{\tL_{1}+\tL_{2}}(\De)\otimes I_{\cW})
U_{2}U_{1}]
\eeqa
where 
$U_{1}=U\otimes I_{\cW}$ and 
$U_{2}=I_{\cH}\otimes e^{-i M\otimes\hat{p}/\hbar}$.
By assumption, we have $[U_{1}U_{2},(\tL_{1}+\tL_{2})\otimes
I_{\cW}]=0$,
so that we have 
\beqa
\et(\tL_{1}+\tL_{2},T_{0},\rh\otimes\si)=0.
\eeqa
Thus, from Theorem \ref{th:URND} we have
\beqa
\ep(\tA,\Pi_{0},\rh\otimes\si)\De (\tL_{1}+\tL_{2})
\ge \frac{1}{2}|\bracket{[\tA,\tL_{1}+\tL_{2}]}|,
\eeqa
where $\tA=A\otimes I_{\cK}$.
We have $\bracket{[\tA,\tL_{1}+\tL_{2}]}=\bracket{[A,L_{1}]}$
and $[\De (\tL_{1}+\tL_{2})]^{2}
=(\De L_{1})^{2}+(\De L_{2})^{2}$.
Thus, we have
\beqa\label{eq:d}
\ep(\tA,\Pi_{0},\rh\otimes\si)^{2}
\ge\frac{|\bracket{[A,L_{1}]}|^{2}}
{4(\De L_{1})^{2}+4(\De L_{2})^{2}}
\eeqa
Let $\rh_{0}=\rh\otimes\si\otimes\ket{\xi}\bra{\xi}$.
Then, we have
\beqas
\lefteqn{
\|U_{1}^{\da}
U_{2}^{\da}
(I_{\cH}\otimes I_{\cK} \otimes \hat{q})
U_{2}U_{1}\sqrt{\rh_{0}} 
-
U_{1}^{\da}
(I_{\cH} \otimes M \otimes I_{\cW})
U_{1}\sqrt{\rh_{0}}\|_{HS}
}\quad\\
&=&
\|U_{1}^{\da}
(I_{\cH}\otimes M\otimes I_{\cW}+
I_{\cH}\otimes I_{\cK} \otimes
\hat{q})U_{1}\sqrt{\rh_{0}}
-
U_{1}^{\da}(I_{\cH} \otimes
 M \otimes I_{\cW})U_{1}\sqrt{\rh_{0}}\|_{HS}\\
&=&
\|U_{1}^{\da}
(I_{\cH}\otimes I_{\cK} \otimes
\hat{q})U_{1}\sqrt{\rh_{0}}\|_{HS}\\
&=&\bracket{\xi|\hat{q}^{2}|\xi}^{1/2}\\
&<&\al.
\eeqas
We also have
\beqas
\lefteqn{
\|U_{1}^{\da}(I_{\cH} \otimes
 M \otimes I_{\cW})U_{1}\sqrt{\rh_{0}}
-
(A\otimes I_{\cK}\otimes I_{\cW})\sqrt{\rh_{0}}\|_{HS}}\quad\\
&=&
\|U^{\da}(I_{\cH} \otimes
 M)U\sqrt{\rh\otimes\si}
-(A\otimes I_{\cK})\sqrt{\rh\otimes\si}\|_{HS}\\
&=&\ep(A).
\eeqas
It follows that we have
\beqas
\lefteqn{
\ep(A\otimes I_{\cK},\Pi_{0},\rh\otimes\si)
}\quad\\
&=&
\|U_{1}^{\da}U_{2}^{\da}(I_{\cH}\otimes I_{\cK} \otimes
\hat{q})U_{2}U_{1}\sqrt{\rh_{0}}
-
(A\otimes I_{\cK}\otimes I_{\cW})\sqrt{\rh_{0}}\|_{HS}\\
&\le&
\|U^{\da}(I_{\cH} \otimes
 M)U\sqrt{\rh\otimes\si}
-(A\otimes I_{\cK})\sqrt{\rh\otimes\si}\|_{HS}\\
& &
{}+\|U_{1}^{\da}
U_{2}^{\da}
(I_{\cH}\otimes I_{\cK} \otimes \hat{q})
U_{2}U_{1}\sqrt{\rh_{0}} 
+
U_{1}^{\da}
(I_{\cH} \otimes M \otimes I_{\cW})
U_{1}\sqrt{\rh_{0}}\|_{HS}
\\
&<&\ep(A)+\al.
\eeqas
Since $\al$ is arbitrary, we have
\beqa\label{eq:e}
\ep(A)\ge
\ep(A\otimes I_{\cK},\Pi_{0},\rh\otimes\si).
\eeqa
Therefore, the assertion follows from Eqs.~\eq{d} and \eq{e}.
\end{Proof}

By the above, the lower bound of the noise decreases 
with the increase of 
the uncertainty of the conserved quantity in the apparatus,
and if the apparatus is macroscopic, the bound can be negligible.

\section{Operational Decoherence in Quantum Logic Gates}

The current theory of fault-tolerant quantum computing 
suggests that the most formidable obstacle for realizing a scalable 
quantum computer is the demand for the high operation precision
for each quantum logic gate, rather than the environment induced
decoherence on quantum memories.
One of the main achievements of this field is
the threshold theorem stating that provided the noise in individual
quantum gates is below a  certain threshold, it is possible to
efficiently perform an arbitrarily large quantum
computing.\cite{NC00} However, the threshold 
is rather demanding.
The current theory 
demands the ``threshold" error probability $10^{-5}$--$10^{-6}$ 
for each quantum gate.
Thus, the fundamental problem turns to whether there is any 
fundamental limit for implementing quantum gates.

In most of current proposals for implementing quantum computing,  
a component of spin of a spin 1/2 system is chosen 
as the computational basis
for the feasibility of initialization and read-out.
For this choice of the computational basis, 
it has been shown\cite{02CQC}
that the angular momentum conservation law limits the accuracy of 
quantum logic operations based on estimating
the unavoidable noise in CNOT gate.
Here, we shall consider the accuracy of implementing 
Hadamard gates,
which are essential components for quantum Fourier transforms
in Shor's algorithm.
In order to implement a quantum circuit for Shor's algorithm
on $L$ bit numbers, we need at least $O(L \log L)$ elementary 
gates in quantum Fourier transform without error correction, 
so that the required  error probability for each Hadamard
gates is below $1/O(L \log L)$ in average. 
This suggests that the accuracy of Hadamard gate is indeed a 
demanding factor in implementing Shor's algorithm.
In what follows, we shall show that Hadamard gates are no easier
to implement under the angular momentum conservation law
than CNOT gates. 

Let $\Q$ be a spin 1/2 system as a qubit with computational basis
$\{\ket{0},\ket{1}\}$ encoded by 
$S_{z}=(\hbar/2)(\ket{0}\bra{0}-\ket{1}\bra{1})$, 
where $S_{i}$ is the $i$ component of spin for $i=x,y,z$.
Let $H=2^{-1/2}(\ket{0}\bra{0}+\ket{1}\bra{0}+\ket{0}\bra{1}-\ket{1}\bra{1})$
be the Hadamard gate $\Q$.

Let $\al=(U,\ket{\xi})$ be a {\em physical implementation} 
of $H$ defined
by a unitary operator $U$ on the system $\bQ+\bA$, where $\bA$ is
a quantum system called the {\em ancilla}, and a state vector
$\ket{\xi}$ of the ancilla, in which the ancilla is prepared
at the time at which $U$ is turned on.  
The implementation $\al=(U,\ket{\xi})$ defines a 
trace-preserving quantum operation $\cE_{\al}$ by
\beq
\cE_{\al}(\rh)=\Tr_{\bA}[U(\rh\otimes\ket{\xi}\bra{\xi})U^{\da}]
\eeq
for any density operator $\rh$ of the system $\bQ$, 
where $\Tr_{\bA}$ stands for the partial trace over the system $\bA$.
On the other hand, the gate $H$ defines the trace-preserving
quantum operation ${\rm ad}H$ by 
\beq
{\rm ad}H(\rh)=H\rh H^{\da}
\eeq
for any density operator $\rh$ of the system $\bQ$.

How successful the implementation $(U,\ket{\xi})$
has been is most appropriately measured by the 
{\em completely bounded (CB) distance}
between two operations $\cE_{\al}$ and ${\rm ad}H$ defined by
\beql{CB-distance}
D_{CB}(\cE_{\al},H)=\sup_{n,\rh}
D(\cE_{\al}\otimes {\rm id}_{n}(\rh),{\rm ad}H\otimes {\rm id}_{n}(\rh)),
\eeq
where $n$ runs over positive integers, ${\rm id}_{n}$ 
is the identity operation
on an $n$-level system $\bS_{n}$,  $\rh$ runs over density operators
of the system $\bQ+\bS_{n}$, and $D(\si_{1},\si_{2})$ stands 
for the trace distance\cite{NC00} of two states $\si_{1}$ and
$\si_{2}$. Since the trace distance of the above two states can be interpreted
as an achievable upper bound on the so-called total variation distance of two
probability distributions arising from measurements performed on the two
output states of the corresponding gates,\cite{NC00}   we interpret
$D_{CB}(\cE_{\al},H)$ as the worst error probability of
operation $\cE_{\al}$ in simulating the gate $H$ on any 
input state of any circuit including those two gates.
We shall call 
$D_{CB}(\cE_{\al},H)$ the {\em gate error probability} of the
implementation $\al$ of the gate $H$. 

Another measure, which is more tractable in computations,
is the {\em gate fidelity}\cite{NC00} defined by
\beq
F(\cE_{\al},H)
=\inf_{\ket{\ps}}F(\ps)
\eeq
where $\ket{\ps}$ varies over all state vectors of $\bQ$, and 
$F(\ps)$
is the fidelity of two states 
$H\ket{\ps}$ and $\cE_{\al}(\ket{\ps}\bra{\ps})$ given by
\beql{fidelity}
F(\ps)=
\bracket{\ps|H^{\da}\cE_{\al}(\ket{\ps}\bra{\ps})H|\ps}^{1/2}.
\eeq
By the relation\cite{NC00} 
\beq
1-F(\cE_{\al},H )^{2}\le D_{CB}(\cE_{\al},H),
\eeq
any lower bound of $1-F(\cE_{\al},H )^{2}$ gives a
lower bound of the gate error probability.
The operator $U$ and the operation $\cE_{\al}$ is generally
described by the following actions on computational basis states
\beqa
U\ket{a}\ket{\xi}&=&\sum_{b=0}^{1}\ket{b}\ket{E^{a}_{b}}\\
\cE_{\al}(\ket{a}\bra{a'})
&=&\sum_{b,b'=0}^{1}\ket{b}
\bracket{E^{a'}_{b'}|E^{a}_{b}}\bra{b'}
\eeqa
for $a,a'=0,1$, where $\ket{E^{a}_{b}}$ is not
necessarily normalized.
It follows that the fidelity is given by 
\beqa
F(\ket{0})^{2}&=&\frac{1}{2}\|\ket{E^{0}_{0}}+\ket{E^{0}_{1}}\|^{2}
=1-\frac{1}{2}\|\ket{E^{0}_{0}}-\ket{E^{0}_{1}}\|^{2},
\label{eq:fidelity-values-1}\\
F(\ket{1})^{2}&=&\frac{1}{2}\|\ket{E^{1}_{0}}-\ket{E^{1}_{1}}\|^{2}
=1-\frac{1}{2}\|\ket{E^{1}_{0}}+\ket{E^{1}_{1}}\|^{2}.
\label{eq:fidelity-values-2}
\eeqa

We consider implementations $(U,\ket{\xi})$ such that
$U$ satisfies the angular momentum conservation law.
For simplicity, we only assume that the $x$ component of the 
total angular momentum is conserved, i.e, 
\beqa
[U,\tS_{x}+\tL_{x}]=0,
\eeqa
where $L_{x}$ is the $x$ component 
of the total angular momentum of the ancilla. 

Now, we consider the following process of 
measuring the operator $S_{z}$ of 
$\Q$: (i) to operate $U$ on $\Q+\bA$, and 
(ii) to measure $S_{x}$ of $\Q$ by a projective measurement.
Since $S_{z}=H^{\dagger}S_{x}H$, 
if $U=H$ the above process would 
measure $S_{z}$ precisely.  
Since each step does not disturb $\tS_{x}+\tL_{x}$,
we can apply \Eq{quantitative-WAY} to this measurement.
Precisely, we consider the instrument $\cI$ for the system
$\cQ+\cA$ defined by
\beq
\cI\{a\}\rh=E^{\tS_{x}}\{a\}U\rh U^{\da}
E^{\tS_{x}}\{a\}
\eeq
for any state $\rh$ of $\cQ+\cA$, where $a=\pm \hbar/2$.
Then, the nonselective operation $T$ of $\cI$ satisfies
\beqa
T[(\tS_{x}+\tL_{x})^{n}]&=&\sum_{a=\pm \hbar/2}U^{\da}
E^{\tS_{x}}\{a\}(\tS_{x}+\tL_{x})^{n}E^{\tS_{x}}\{a\}U\\
&=&
(\tS_{x}+\tL_{x})^{n}.
\eeqa
Thus, we have 
\beqa
\et(\tS_{x}+\tL_{x},T,\rh)=0.
\eeqa
Thus, the POVM $\Pi$ of $\cI$ satisfies
\beqa
\ep(\tS_{z},\Pi,\rh)\ge\frac{|\bracket{[\tS_{x},\tS_{z}]}|}
{2\De(\tS_{x}+\tL_{x})}.
\eeqa
Let $\rh=\ket{\ps}\bra{\ps}\otimes\ket{\xi}\bra{\xi}$.
Then, we have 
$\bracket{[\tS_{x},\tS_{z}]}=\bracket{[S_{x},S_{z}]}$,
$
[\De(\tS_{x}+\tL_{x})]^{2}=
(\De S_{x})^{2}+(\De L_{x})^{2}$, 
and 
$\ep(\tS_{z},\Pi,\ps\otimes\xi)^{2}=\ep(S_{z},\Pi_{0},\ps)$,
where
$\Pi_{0}\{a\}=\Tr_{\bA}[\Pi\{a\}(I\otimes\ket{\xi}\bra{\xi}]$
for $a=\pm \hbar/2$,
so that we have
\beqa\label{eq:LB}
\ep(S_{z})^{2}\ge 
\frac{|\bracket{[S_{z},S_{x}]}|^{2}}{4(\De S_{x})^{2}+4(\De
L_{x})^{2}},
\eeqa
where $\ep(S_{z})=\ep(S_{z},\Pi_{0},\ps)$.
Now, we have
\beqas
\ep(S_{z})^{2}
&=&
\bracket{\ps|
O^{(2)}(\Pi_{0})-O(\Pi_{0})S_{z}-S_{z}O(\Pi_{0})+S_{z}^{2}
|\ps}\\
&=&
\bracket{\ps\otimes\xi|(U^{\da}\tS_{x}U-\tS_{z})^{2}|\ps\otimes\xi}\\
&=&
\|\tS_{x}U\ket{\ps\otimes\xi}-U\tS_{z}\ket{\ps\otimes\xi}\|^{2}\\
&=&
\frac{\hbar^{2}}{2}|\bracket{0|\ps}|^{2}\|\ket{E^{0}_{0}}-\ket{E^{0}_{1}}\|^{2}
+
\frac{\hbar^{2}}{2}|\bracket{1|\ps}|^{2}\|\ket{E^{1}_{0}}+\ket{E^{1}_{1}}\|^{2}.
\eeqas
Thus, from \Eq{fidelity-values-1} and \Eq{fidelity-values-2},
we have
\beqa
\frac{\ep(S_{z})^{2}}{\hbar^{2}}
=1-|\bracket{0|\ps}|^{2}F(\ket{0})^{2}+|\bracket{1|\ps}|^{2}F(\ket{1})^{2}
\eeqa
It follows that we have
\beqa
1-F(\cE_{\al},H)^{2}
&=&1-\inf_{\ket{\ps}}F(\ps)^{2}\\
&\ge& 1-|\bracket{0|\ps}|^{2}F(\ket{0})^{2}+|\bracket{1|\ps}|^{2}F(\ket{1})^{2}\\
&=&\frac{\ep(S_{z})^{2}}{\hbar^{2}}
\eeqa

For the input state 
$\ps=(\ket{0}+i\ket{1})/\sqrt{2}=\ket{S_{y}=\hbar/2}$,
the numerator $|\bracket{[S_{z},S_{x}]}|^{2}$ of the lower bound
\eq{LB} is maximized as
\beqa\displaystyle\label{eq:bound}
1-F(\cE_{\al},H)^{2}\ge\frac{\ep(S_{z})^{2}}{\hbar^{2}}
\ge\frac{1}{4+4 (2\De L_{x}/\hbar)^{2}}.
\eeqa
Similar result on CNOT gates were previously obtained in
Ref.~\onlinecite{02CQC} (see also, Ref.~\onlinecite{Lid03,02CQCReply}).
Here, we have shown that the Hadamard gate, a single qubit
gate, has the  unavoidable error probability equivalent to that for the
CNOT.

In the following, we shall interpret the above relation for
bosonic control systems and fermionic control systems separately.
In current proposals, the external electromagnetic field prepared by
laser  beam is considered to be a feasible candidate for the controller
$\bA$ to be coupled with the computational qubits $\bQ$ via the
dipole interaction.\cite{NC00}  In this case, the ancilla state 
$\ket{\xi}$ is considered to be a coherent state, for which we have 
$(\De N)^{2}=\bracket{\xi|N|\xi}=\bracket{N}$, 
where $N$ is the number operator.  
We assume that the beam propagates to the
$x$-direction with right-hand-circular polarization.  Then, we
have $L_{x}=\hbar N$,  and hence
\beq
(2\De L_{x}/\hbar)^{2}=
(2 \De N)^{2}=4\bracket{N}
\eeq
Thus, from \Eq{bound} we have
\beqa
 1-F(\cE_{\al},H)^{2}\ge\frac{1}{4+16\bracket{N}}.
\eeqa
Thus, we cannot implement Hadamard gates within the error
probability $(4+16\bracket{N})^{-1}$ on a qubit represented 
by a spin component of a spin 1/2 system controlled by the dipole 
interaction with external electromagnetic field with average 
photon number $\bracket{N}$.
Enk and Kimble\cite{EK02} and Gea-Banacloche\cite{Ban02} 
also showed that there is unavoidable error probability in this case
inversely proportional to the average strength of the external field 
by calculations with the model obtained by rotating wave 
approximation.
Here, we have shown the same result only from the angular
momentum conservation law.

We now assume that the ancilla $\bA$ comprises $n$ spin 1/2
systems.
Then, we have
\beq
\De L_{x}\le \|L_{x}\|=\frac{n\hbar}{2}.
\eeq
Thus, we have the following lower bound of the gate
error probability
\beq
 1-F(\cE_{\al},H)^{2}\ge\frac{1}{4+4n^{2}}.
\eeq
Thus, it has been proven that if the computational basis is 
represented by the $z$-component of spin, we cannot implement
Hadamard gates within the error probability 
$(4+4n^{2})^{-1}$ with $n$ qubit ancilla by rotationally invariant
interactions such as the Heisenberg exchange interaction.
Thus,  for the error probability $\sim 10^{-5}$,
we need the ancilla consisting of at least $\sim 100$ physical qubits.
This result shows a drastic contrast with the  new 
universal encoding of the computational qubit
recently proposed by DiVincenzo et al.\cite{DBKBW00}.
In their encoding, each computational qubit is encoded into three
physical qubits,  instead of one spin 1/2 system, and
they showed that any quantum gates for $n$ logical qubits are
implemented with arbitrary accuracy  by rotationally invariant
interactions on $3n$ physical qubits, so that Hadamard gates are
implemented only on three physical qubits with required accuracy.

In the above discussion, we have assumed that the control 
system can be prepared in an entangled state.  However, 
it is also interesting to estimate the error in the case where
we can prepare the control system only in a separable
state.
In this case, we have
\beq
(\De L_{x})^{2}\le \sum_{j=1}^{n}(\De S_{x}^{(j)})^{2}
\le n\|S_{x}\|^{2}=\frac{n\hbar^{2}}{4},
\eeq
where $ S_{x}^{(j)}$ is the spin component of the $j$th
ancilla qubit so that $L_{x}=\sum_{j=1}^{n}S_{x}^{(j)}$.
Thus, we have the following lower bound of the gate
error probability
\beql{bound2}
1-F(\cE_{\al},H)^{2}
\ge \frac{\ep(S_{z})^{2}}{\hbar^{2}}
\ge \frac{1}{4+4n}.
\eeq
Thus, the error probability is lower bounded by $(4+4n)^{-1}$,
and hence the achievable error can be considered to be
inversely proportional to $4n^{2}$ for entangled control system
but $4n$ for separable control system.  Note that even if
the ancilla is in a separable mixed state, the relation
$(4+4n)^{-1} \le \ep(S_{z})^{2}/\hbar^{2}$
still holds, since $\ep(S_{z})^{2}$ is an affine function
of the ancilla state.

If  the field is in a number state $\ket{n}$,
then 
\beq
(2\De L_{x}/\hbar)^{2}=
(2 \De N)^{2}=0,
\eeq
so that we have
\beqa
\frac{\ep(S_{z})^{2}}{\hbar^{2}}
\ge\frac{1}{4}.
\eeqa
Thus, if  the field state is a mixture
of number states such as the thermal state, i.e., 
$\si=\sum_{n}p_{n}\ket{n}\bra{n}$, we have also
the lower bound $\ep(S_{z})^{2}/\hbar^{2}
\ge{1}/{4}$.
Thus, it seriously matters whether the control field is 
really in a coherent state or a mixture of number states.

\section{Conclusions}

The notion of quantum instruments is formalized by normalized
completely positive map valued measures to represent statistical
equivalence  classes of all the possible quantum measurements.
Universally valid operational uncertainty relations are established
to set a precision limit for any instrument given a disturbance
constraint.  The Heisenberg relation on the lower bound for the
product of the root-mean-square noise and disturbance is
derived for those instruments with uncorrelated noise
and disturbance from a universal uncertainty relation.  A new
precision bound for nondisturbing instruments follows immediately
from the universal uncertainty relation and leads to a quantitative 
generalization of the Wigner-Araki-Yanase theorem on the
precision limit of measurements under conservation laws.  
Applying this,  a rigorous lower bound is obtained for the gate
error probability of any physical realizations of the Hadamard gate
under the constraint that the computational basis is represented by
a component of spin of a spin 1/2 system, and that physical
implementation obeys the angular momentum conservation law.
The lower bound is shown to be $1/(4+16\bracket{N})$ 
for the external control field with average photon number
$\bracket{N}$ in a coherent state, whereas it amounts to
$1/4$ for the field in the thermal state.
For fermionic control, the lower bound is 
$1/(4+4n^{2})$ for $n$ qubit ancilla in an
entangled state, and $1/(4+4n)$ in a separable states.
All of these lower bounds have been obtained from rigorous
calculations without any approximations under the sole assumption
of the angular momentum conservation law.
Physical significance of those fundamental lower bounds deserve
further investigations and will be discussed elsewhere.

\section*{Acknowledgements}

This work was supported in part by the 
SCOPE project of the MPHPT of Japan,  by the CREST
project of the JST, and by the Grant-in-Aid for Scientific Research of
the JSPS.

\end{document}